# Scanning Tunneling Microscopy Study of Epitaxial $Fe_3GeTe_2$ Monolayers on $Bi_2Te_3$

*Brad M. Goff, Wenyi Zhou, Alexander J. Bishop, Ryan Bailey-Crandell, Katherine Robinson, Roland K. Kawakami, Jay A. Gupta\**

*Department of Physics, The Ohio State University, Columbus, OH 43210, USA*

**Abstract**

Introducing magnetism to the surface state of topological insulators, such as $Bi_2Te_3$, can lead to a variety of interesting phenomena. We use scanning tunneling microscopy (STM) to study a single quintuple layer (QL) of the van der Waals magnet $Fe_3GeTe_2$ (FGT) that is grown on $Bi_2Te_3$ via molecular beam epitaxy. STM topographic images show that the FGT grows as free-standing islands on $Bi_2Te_3$ and outwards from $Bi_2Te_3$ steps. Atomic resolution imaging shows triangular lattices of 390 ± 10 pm for FGT and 430 ± 10 pm for $Bi_2Te_3$, consistent with the respective bulk crystals. A moiré pattern is observed on FGT regions with a periodicity of 4.3 ± 0.4 nm that can be attributed solely to this lattice mismatch and thus indicates zero rotational misalignment. While most of the surface is covered by a single QL of the FGT, there are small double QL regions, as well as regions with distinct chemical terminations due to an incomplete QL. The most common partial QL surface termination is the FeGe layer, in which the top two atomic layers are missing. This termination has a distinctive electronic structure and a $(\sqrt{3} \times \sqrt{3})R30°$ reconstruction overlaid on the moiré pattern in STM images. Magnetic circular dichroism (MCD) measurements confirm these thin FGT films are ferromagnetic with $T_C$ ~190 K.

\* Corresponding author: gupta.208@osu.edu

**Introduction**

Introducing magnetism to topological insulators (TIs) breaks time-reversal symmetry and leads to a variety of interesting phenomena such as magnetoelectric effects [1, 2], Weyl semimetal phases [3], efficient coupling of quantum spin defects [4], and the quantum anomalous Hall effect (QAHE) [5, 6]. Some approaches to create magnetic TIs include: doping with magnetic impurities, proximity to magnetic layers, and intrinsic magnetism [5–7]. The QAHE has been observed in each of these cases, but only at temperatures much lower than the Curie temperature. The physics underlying this discrepancy is not well understood but may be due to disorder associated with random doping or interfaces [8]. Integration of TIs with van der Waals (vdW) ferromagnets offers a route to minimize interfacial disorder, as the vdW gap eliminates dangling bonds and facilitates epitaxial growth of materials with disparate lattices. Among the growing family of vdW ferromagnets, $Fe_3GeTe_2$ (FGT) is particularly attractive to couple to TIs such as $Bi_2Te_3$ due to its relatively similar layered structure and large bulk Curie temperature ($T_C$ ~ 220 K [9] ) which persists down to the monolayer limit ($T_C$~ 130 K [10]) and may even further increase up to room temperature in FGT/$Bi_2Te_3$ heterostructures [11, 12]. Incorporation of FGT into heterostructures has also yielded skyrmion spin textures [13–15] and spin-orbit torque switching [16–18], which is promising for future device applications. Additionally, the lattice mismatch between FGT and $Bi_2Te_3$ produces a moiré pattern which has the potential to be used as a trapping site for skyrmions [19].

Here we report the first scanning tunneling microscopy (STM) studies of single quintuple layer (QL) FGT, grown on $Bi_2Te_3$ via molecular beam epitaxy. Topographic images show that the FGT grows both as isolated islands on $Bi_2Te_3$ terraces, and outward from $Bi_2Te_3$ steps. Atomic resolution imaging shows triangular lattices of 390 ± 10 pm for FGT and 430 ± 10 pm for $Bi_2Te_3$, both consistent with their respective bulk crystals. The FGT has a moiré pattern with a periodicity of 4.3 ± 0.4 nm attributed solely to the lattice mismatch showing that the FGT is rotationally aligned to the $Bi_2Te_3$. While most of the surface is covered by 1 QL FGT, smaller regions of 2 QL FGT and fractional QL terminations are identified. Atomic resolution imaging and tunneling spectroscopy are used to track changes in lattice and electronic structure with varying thickness in this regime. While 1 QL and 2 QL FGT are found to be nearly identical, the FeGe partial termination exhibits a distinct surface reconstruction and electronic structure. In addition, we report reflection magnetic circular dichroism (MCD) measurements confirming that these FGT thin films are ferromagnetic with a Curie Temperature of ~190 K. These studies establish the FGT/ $Bi_2Te_3$ system as a useful model for probing the coupling of magnetism and topological states with minimal doping and interfacial disorder.

**Experimental methods**

The MBE films were deposited in a Veeco 930 MBE system with a base pressure of $1\times10^{-10}$ mbar. The c-sapphire substrates were annealed in air at 1000 °C for 3 hours before depositing an elemental Pt on one corner of the substrate. The substrate was then mounted on a standard omicron-style paddle and a small Ta foil clip was put in contact with the Pt corner and spot welded to the paddle. This addition of Pt and Ta foil serves as an electrical conduction path between the deposited films and the paddle and is necessary to facilitate tunneling between the MBE film and STM tip. After mounting, the substrate was annealed *in situ* at 800 °C for one hour and cooled to 250 °C for the growth of $Bi_2Te_3$ films through a two-step process. The first three $Bi_2Te_3$ QLs were grown at a substrate temperature of 250 °C. The Bi shutter was then closed, stopping any further growth of material and the substrate temperature was increased to 295 °C and annealed in Te for 15 minutes. Next, the Bi shutter was opened again, continuing the growth of $Bi_2Te_3$ for an additional 15 QLs. Flux ratios of Bi (99.998%, Alfa Aesar) to Te (99.9999%, United Mineral Corp) were maintained at ~1:15 for each stage of the $Bi_2Te_3$ growth as measured by a beam flux monitor (BFM). This process resulted in a high-quality $Bi_2Te_3$ film with an overall thickness of 18 nm. Immediately following the Bi$_2$Te$_3$ growth the substrate temperature is again raised, while under Te flux, to 325 °C for the growth of the FGT films. This was accomplished by co-depositing Fe (99.99%, Alfa Aesar), Ge (99.9999%, Alfa Aesar), and Te with a calibrated atomic flux ratio of 3:1:20, as measured by BFM. These growth conditions match our earlier work of FGT growth on Ge(111) substrates [20]. The deposition rate was calibrated by fringes in the X-ray reflectivity of a ~24 nm thick FGT film. This gives a growth rate of 1 monolayer equivalent (MLe) every ~90 seconds.

FGT films (1 MLe and 1.5 MLe) were transferred from the growth chamber to the STM chamber using a Ferrovac UHV suitcase (P < 10$^{-9}$ mbar) and STM measurements were performed without any additional surface preparation. STM measurements were performed under ultra-high vacuum (1x10$^{-10}$ mbar) in a CreaTec LT-STM at 5 K. To potentially probe magnetism in the samples via spin polarized STM (SP-STM) [21], electrochemically etched Cr and Ni tips were used [22, 23]. The tips were prepared for imaging by Ar$^+$ ion sputtering and field emission. Scanning tunneling spectroscopy (STS) measurements were performed using lock-in detection of the differential conductance, dI/dV, by adding an AC bias modulation of 50-100 mV$_{rms}$ at 877 Hz and with the feedback loop turned off to keep the tip-sample separation constant as the bias voltage is swept. We also acquired simultaneous topographic and dI/dV maps at fixed voltage by using the same lock-in technique with the feedback loop on. Analysis of STM images was performed using Gwyddion [24].

Out-of-plane magnetic hysteresis loops were measured using MCD on 1.5 MLe FGT films, capped with a transparent ~5nm CaF$_2$ layer as the MCD cryostat does not have vacuum transfer capabilities. The MCD measurement was performed in high vacuum (1x10$^{-8}$ mbar) in a closed-cycle cryostat. A continuous wave 532 nm laser was focused to a ~20 um spot on the sample, with modulation between right and left circular polarization at 50 kHz using a photoelastic modulator (PEM). The intensity of the reflected beam was detected using a photodiode and lock-in amplifier which was used to calculate the MCD ratio ($\frac{I_{RCP}-I_{LCP}}{I_{RCP}+I_{LCP}}$), which is proportional to the magnetization.

**Results and Discussion**

Figure 1a shows a schematic of the FGT/Bi$_2$Te$_3$ heterostructure. Fe$_3$GeTe$_2$ and Bi$_2$Te$_3$ have similar QL structures, with a vdW gap between Te layers. Within each QL, FGT has a stacking order of Te, Fe, FeGe, Fe and Te layers, each arranged in a triangular lattice and held together via covalent bonds. The Te layer is the energetically preferred surface termination, both in exfoliated crystals and in epitaxial thin films, though surfaces can terminate on different atomic layers, as seen in other vdW materials such as MnBi$_2$Se$_4$ [25], Bi$_2$Se$_3$, and Bi$_2$Te$_3$ [26]. In Fig. 1b, we show a typical large-scale topographic STM image of the FGT/Bi$_2$Te$_3$ surface, colorized to highlight the distinct terminations indicated by the labeling. Regions of exposed Bi$_2$Te$_3$ are identified by comparison with STM imaging of a Bi$_2$Te$_3$ control sample (c.f. supporting information) and show a characteristic adsorbate density and step height. The line profile in Figure 1c shows the changes in apparent height starting from a Bi$_2$Te$_3$ terrace on the left to a neighboring one on the right. The Bi$_2$Te$_3$ terraces are separated by a step height of 1.00 ± 0.03 nm which is consistent with the bulk Bi$_2$Te$_3$ QL spacing. The chosen profile also crosses two qualitatively distinct regions that are assigned to FGT: starting from the left there is a step up of 0.85±0.02 nm to the first 'coplanar' FGT region, which appears to grow outward from the Bi$_2$Te$_3$ step, and there is then an identical step up to a free-standing FGT island on the neighboring terrace. Statistically, we find most of the FGT regions tend to grow from steps, rather than form free-standing islands. STM measurements of the Bi$_2$Te$_3$ control sample showed more triangular-shaped terraces separated by relatively straight step edges, in contrast to the more meandering step edges in Fig. 1b. This suggests that the FGT growth and/or the associated higher growth temperature may cause a reshaping of the $Bi_2Te_3$ steps.

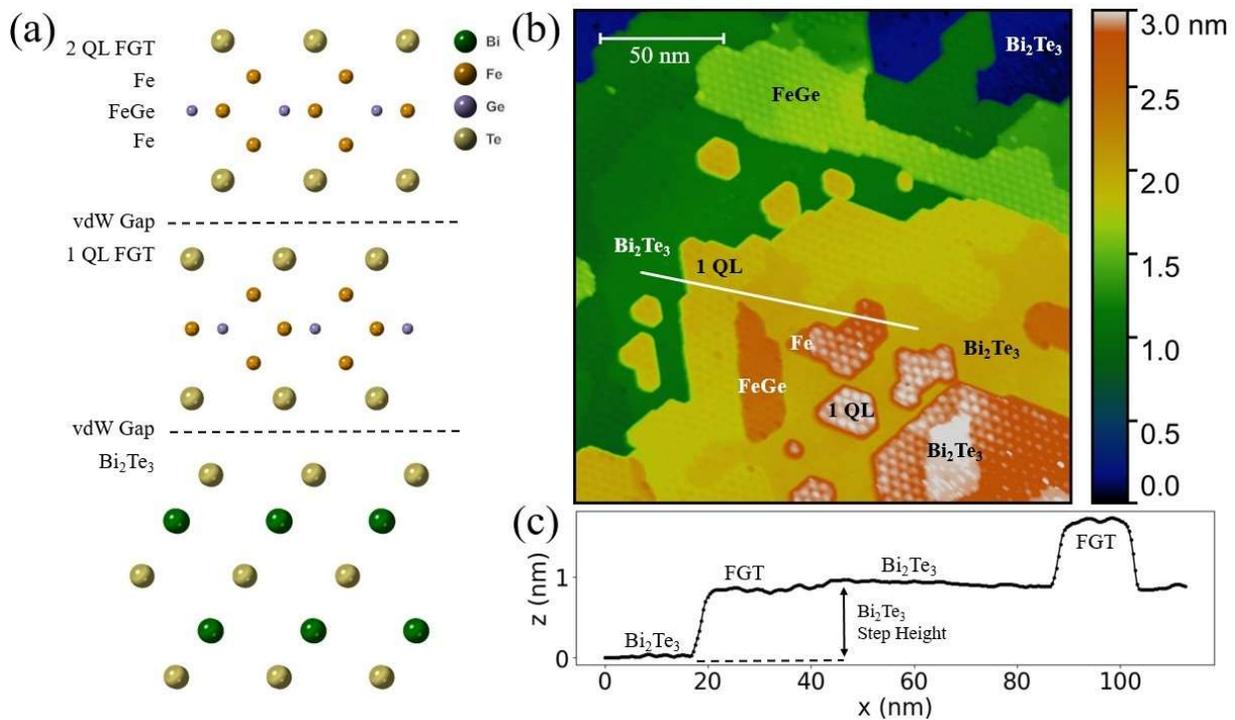

**Fig 1:** FGT / $Bi_2Te_3$ film morphology. (a) Side view of the crystal structure with the observed terminations labeled. (b) Typical topographic STM image (-0.8 V, 0.3 nA), colorized and labeled to highlight the assigned surface terminations (c) The height profile corresponding to the line in (b).

Atomic-resolution STM imaging and STS do not indicate any significant differences between the co-planar and island FGT, suggesting that the additional $Bi_2Te_3$ interface for co-planar FGT doesn't introduce additional strain or otherwise affect the FGT properties. Figure 2a shows one such image of the interface between coplanar FGT and $Bi_2Te_3$. A triangular lattice of bright spots is observed on $Bi_2Te_3$ which we attribute to the terminating Te atoms, similar to prior STM studies of bulk $Bi_2Te_3$ [27]. We attribute the smattering of bright spots in this area to Te adatoms, as we found a similar density of such features in the control $Bi_2Te_3$ sample, and we attribute this to Te-rich growth conditions. A low density of dark spots is also evident in the image, which we attribute to $Te_{Bi}$ antisite defects. This is consistent with the continuity of the Te lattice over the defects (not shown), and previous STM studies [28]. The lower half of the image is a single QL of FGT, growing outward from the $Bi_2Te_3$ step separating the two regions. We attribute the lattice of bright spots on FGT to a moiré pattern, associated with the superposition of the FGT and $Bi_2Te_3$ lattices. The moiré spots are non-uniform, adopting both triangular and more spherical shapes which could reflect a distribution of adatoms and defects in the FGT.

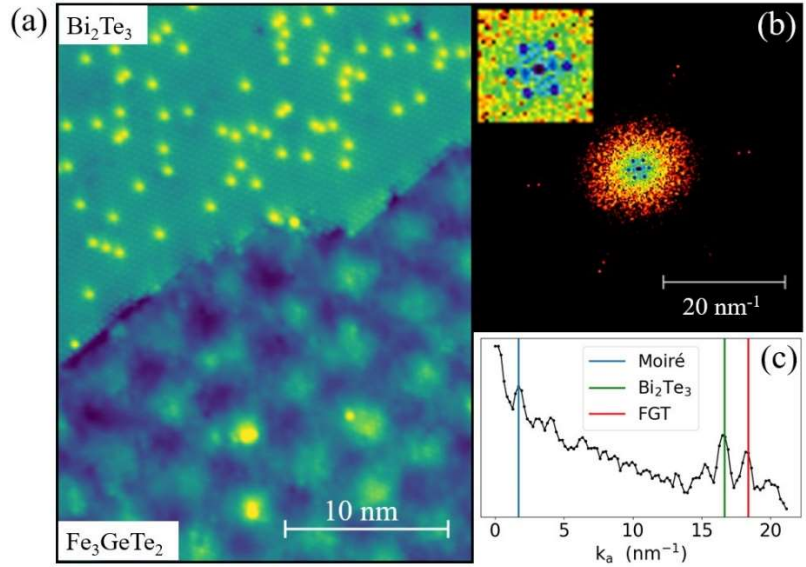

**Fig 2:** FGT and Bi$_2$Te$_3$ interface. (a) Image of FGT and Bi$_2$Te$_3$ atomic lattices (-0.8 V, 0.3 nA). The moiré pattern is visible in the FGT region. (b) The fast-Fourier transformation of (a) with peaks corresponding to the FGT and Bi$_2$Te$_3$ Brillouin zone. The inset shows the peaks corresponding to the moiré periodicity in the center. (c) A linescan along the peaks of (b), used to calculate the FGT, Bi$_2$Te$_3$, and moiré lattice constants of 390 ± 10 pm, 430 ± 10 pm, and 4.3 ± 0.4 nm respectively.

Though not discernible at the magnification presented in the Figure, a triangular atomic lattice is also resolved on FGT and is more readily analyzed in the corresponding Fast Fourier Transform, Fig. 2b. The FFT shows hexagonal spot patterns associated with these three lattice structures, and lattice constants are extracted from a profile of the spot pattern (Fig. 2c). The measured lattice constant of the Bi$_2$Te$_3$ region is 430 ± 10 pm (green line in Fig. 2c), which is consistent with the bulk crystal (438.6 pm) and our diffraction measurements (c.f. supporting information). The FGT atomic lattice produces the outermost spot pattern, and is exactly aligned with the Bi$_2$Te$_3$ lattice spots in the FFT, indicating alignment of the corresponding real-space lattices. From these data we extract an FGT lattice constant of 390 ± 10 pm, which is also within uncertainty of the bulk crystal (399 pm) [9]. This suggests a low degree of interfacial strain, as expected from the weak bonding across the van der Waals gap between the FGT and underlying Bi$_2$Te$_3$. The magnified inset of the central region in the FFT shows the third set of hexagonal spots associated with the moiré lattice, and indicates a spacing of 4.3 ± 0.4 nm. This is consistent with the calculated moiré spacing of $a_m = \left(1/a_{FGT} - 1/a_{Bi2Te}\right)^{-1}$ (for zero rotation angle) using the observed lattice constants ($a_m$= 4.2 nm), or assuming bulk values ($a_m$= 4.47 nm). This again indicates low strain at the interface, as the moiré spacing is a sensitive probe of variations in lattice and/or rotational mismatch. We find similar moiré spacings for all FGT regions imaged, whether coplanar or freestanding.

While the Te-termination of FGT surfaces is energetically preferred, we observed regions with other terminations in our MBE-grown samples. For example, Figure 1b shows a small region labeled 'Fe' which is atomically flat, but only a fractional QL step up from the surrounding terrace, consistent with termination on the Fe layers in Fig. 1a. Figure 3a shows an STM image of the most common of the partial terminations, which we attribute to the FeGe layer that lies within the middle of the FGT QL. A line profile in this region (Fig. 3b) shows a step up from the 1 QL FGT region on the right, to an intermediate terrace in the middle and then to a higher island on the left. The total step height in this case (800 ± 50 pm) is in good agreement with the QL spacing in bulk FGT (816.8 pm) [9], and thus we identify 2 QL FGT as the left terrace. Bright contrast within the 2 QL terrace indicates a moiré pattern with a spacing identical to that observed in the 1 QL regions, with however, only 1/3 of the apparent height modulation amplitude. This suggests the moiré modulation is associated with the FGT/ $Bi_2Te_3$ interface, and that the two QL sheets are aligned in the 2 QL FGT regions.

The central terrace in Figure 3a which we attribute to the FeGe layer is atomically flat and corresponds to a step up of 640 ± 30 pm from the 1 QL FGT region (Fig. 3b). This is ~75% of the QL step height and corresponds well to the van der Waals gap plus the spacing of Te-Fe-FeGe layers. We also observe a moiré lattice on the FeGe regions (e.g. the dark honeycomb lattice in Fig. 3a) with the same periodicity as on the 1 QL regions, again reflecting the underlying FGT / $Bi_2Te_3$ interface. Though the moiré periodicity is identical within error on all FeGe regions imaged, there is variation in the appearance of the moiré spots as shown in Fig. 3c, which may be due to underlying defects or small amounts of strain. Figure 3c also shows an atomic lattice with a spacing of 690 ± 20 pm which is larger than that observed on other FGT regions. This lattice is consistent with a $(\sqrt{3} \times \sqrt{3})R30°$ reconstruction of FGT, suggesting that there is some buckling of the FeGe layer when exposed as a surface.

Having identified the range of distinct surfaces from our STM images, we explored variations in electronic structure using a combination of STS and dI/dV mapping. Figure 4a shows a compilation of tunneling spectra taken on the exposed $Bi_2Te_3$, 1 and 2 QL FGT, and the FeGe surfaces. The $Bi_2Te_3$ spectrum is consistent with the previous literature, with a rise near 0 V associated with the $Bi_2Te_3$ conduction band edge, and a rise at ~ -150 mV associated with the bulk valence band edge [26, 27, 29]. Roughly linear dispersion within the bulk gap extrapolates to a Dirac point of ~ -250 mV [29]. The similar position of the Dirac point in our MBE thin films with prior STM studies of bulk crystals suggests that residual doping from impurities or native defects is similar in the two classes of material. On both 1 QL and 2 QL FGT regions, STS show a sharp dip centered at the Fermi level, and a broader occupied-state peak around -300 mV. The Fermi-level dip is also observed on the FeGe termination, and there is a shift in spectral weight among the broad peaks observed in occupied and unoccupied states.

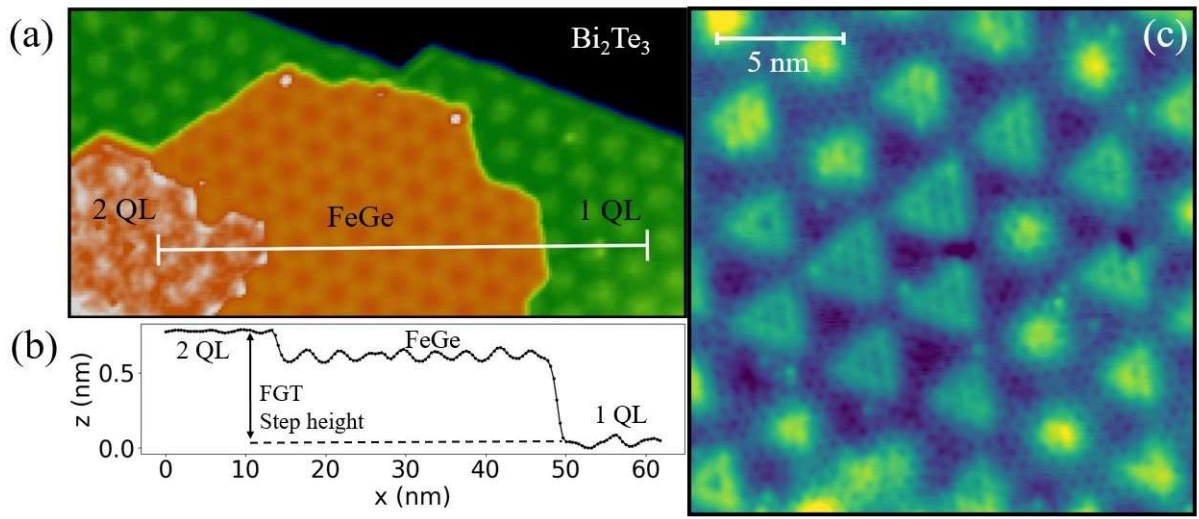

**Fig 3:** STM imaging of different terminations in the FGT/$Bi_2Te_3$ thin films. (a) A topographic image showing FeGe and 2 QL FGT terminations. (-0.8 V, 0.1 nA) (b) The height profile corresponds to the line in (a) and indicates QL and fractional QL step heights. (c) STM image of a second region of FeGe termination showing a mixed moiré pattern and an atomic-like lattice of dark spots attributed to a $(\sqrt{3} \times \sqrt{3})R30°$ surface reconstruction. (1 V, 0.3 nA)

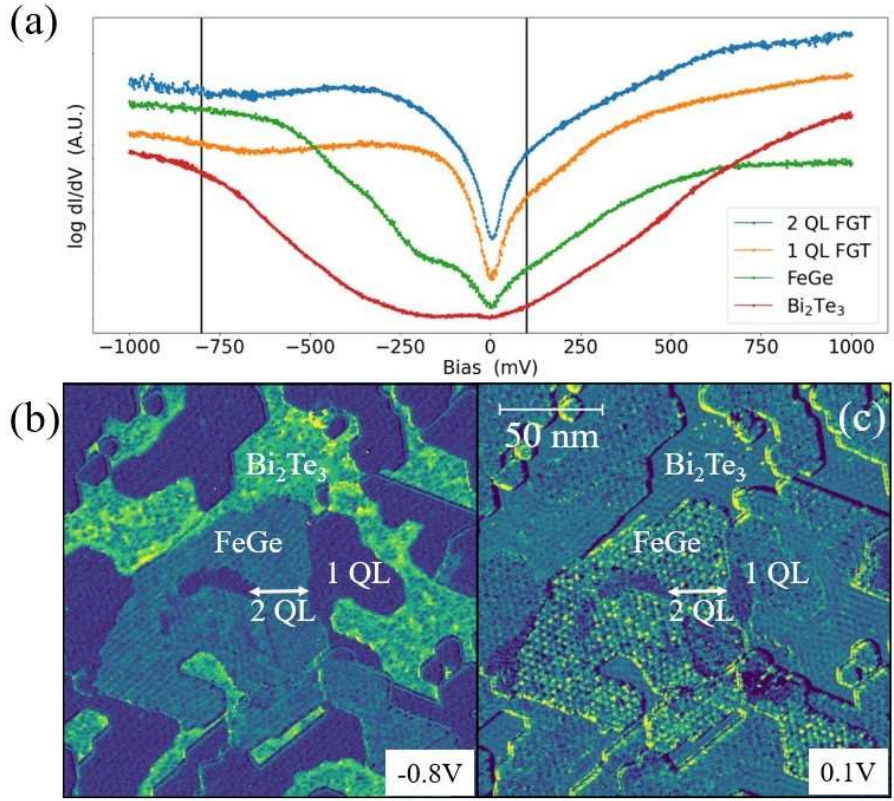

**Fig 4:** Electronic structure of different terminations. (a) STS on $Bi_2Te_3$, 1 QL FGT, 1.6 QL FGT, and 2 QL FGT. The spectra are offset and plotted on a log scale for clarity. Two vertical lines show the bias used for images (b) and (c). (b) dI/dV image at a bias of –0.8 V with different regions labeled. (0.1 nA) 1 and 2 QL FGT cannot be distinguished from this image without the simultaneously taken topographic image. (c) dI/dV image at a bias of 0.1 V with different regions labeled (0.1 nA). The $Bi_2Te_3$ surface state is visible as a standing wave

The similarity of our observed STS on 1, 2 QL FGT with literature reports on bulk FGT [30–32] indicate that there is relatively little change in density of states with layer thickness. This is also consistent with FGT slab density functional theory (DFT) calculations [33]. Though the detailed shape of spectra in STS measurements in the literature vary, the dip near $E_F$ has been attributed to a Kondo resonance [30, 34] or inelastic spin scattering [32]. Broad peaks associated with occupied states in this range have also been reported previously in the prior bulk STS, and temperature-dependent ARPES measurements confirm an association with the ferromagnetic order [30]. DFT calculations don't capture the physics underlying the Fermi-level dip, but do indicate that Fe $d$-states dominate the DOS near $E_F$, consistent with the broader features observed on both the Te- and FeGe-terminated surfaces in Figure 4.

Previous SP-STM studies on bulk FGT indicated contrast in spatial maps of the dI/dV signal, associated with micron-scale spin up/down domains [31, 32]. To look for magnetism in our FGT thin films via SP-STM, we show dI/dV maps taken at – 800 mV (Fig. 4b) and + 100 mV (Fig. 4c), with the layer assignments based on topographic images (not shown). In Figure 4b, $Bi_2Te_3$ images with the brightest contrast, punctuated by Te adatoms which appear as dark point defects. Single QL and 2 QL FGT are largely indistinguishable at this bias, consistent with the similar dI/dV spectroscopy in Fig. 4a. The FeGe-terminated layers are imaged with medium contrast at this bias, and are readily distinguished from the 1 QL and 2 QL FGT regions. In dI/dV maps at 100 mV bias (Fig. 4c), $Bi_2Te_3$ regions image with medium contrast and show a periodic quasiparticle interference pattern [29]. These correspond to Friedel oscillations of the topological surface state electrons, and indicate that the topological surface state persists in these 18 QL $Bi_2Te_3$ thin films, even when the surface is mostly covered with FGT. The brightest contrast at this bias is found on FeGe regions, which also show a strong modulation associated with the moiré lattice. In addition, 1 QL and 2 QL FGT regions can be distinguished, as the 2 QL FGT regions image with darker contrast in the dI/dV images at this bias.

Having characterized the dI/dV contrast associated with the different surfaces in our samples, we repeatedly imaged these areas with varying tip terminations and applied out-of-plane magnetic fields of ±1T to look for the expected domain structures in the FGT regions. These efforts were inconclusive however, and all FGT-related regions presented spatially-uniform contrast in our dI/dV images. We confirm that these films are ferromagnetic by performing MCD measurements. Figure 5a shows the temperature dependence of the out-of-plane hysteresis loops of a ~1.5 MLe FGT on an 8 nm $Bi_2Te_3$ film from 140 K to 200 K. The symmetric part of each hysteresis loop was subtracted to eliminate symmetric drift of signal due to sample drift in a magnetic field. Also, a constant offset has been added to each curve to better compare the hysteresis loops. At 140 K, the MCD exhibits a relatively square hysteresis loop with a coercivity of 500 Oe. Increasing from 140 K to 200 K, the coercivity decreases and the remanent magnetization gets smaller. The magnitude of the MCD signal reduces to zero at 200 K, indicating a Curie temperature ($T_C$) between 180 K and 200 K which is consistent with previous works [11]. To confirm the average surface coverage of this sample, an identical 1.5 MLe FGT sample (but without the capping layer) was imaged in the STM with a bulk Ni tip. The largely uniform dark contrast in the dI/dV map in Fig. 5b indicates most of the sample is covered with 1 QL and 2 QL FGT, with few $Bi_2Te_3$ and FeGe regions compared to the 1 MLe sample. The expected increase in coercivity at 5K suggested by the MCD exceeds our field range in the SP-STM measurements, which could explain the absence of switching or domain structure in our dI/dV images.

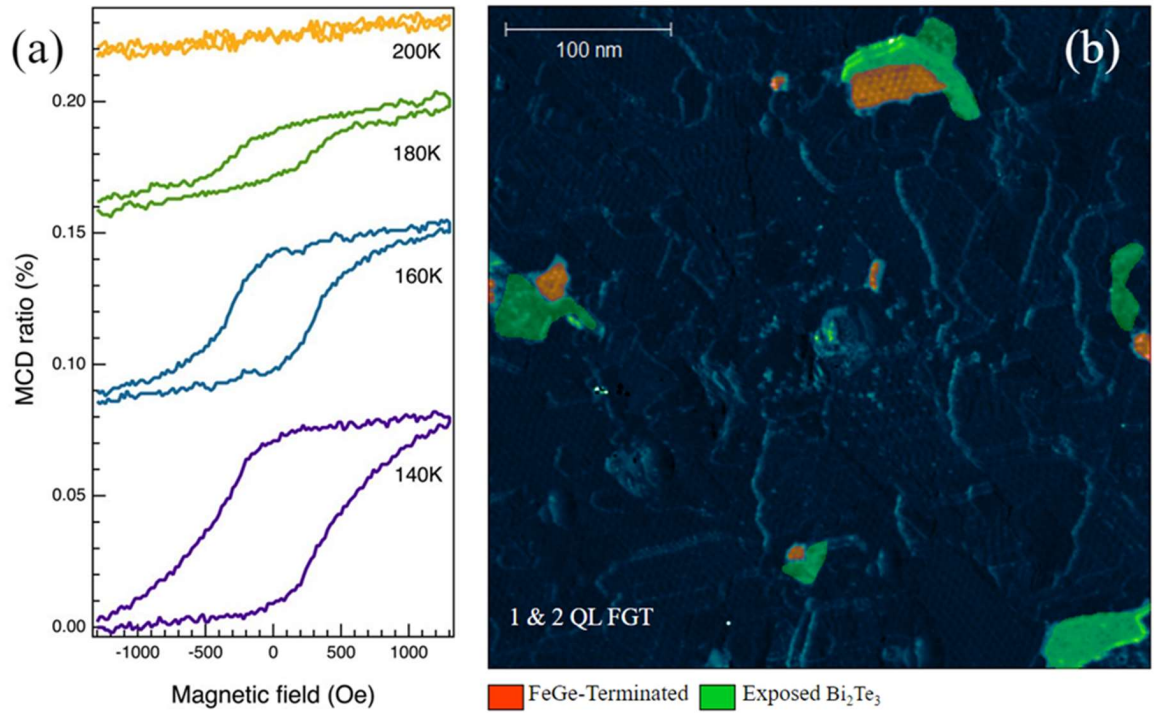

**Fig 5:** Measurements of 1.5 MLe FGT film on 8 MLe Bi2Te3 (a) Temperature dependent out-of-plane hysteresis loops measured by MCD. (b) dI/dV STM image. The FeGe-terminated FGT regions and the exposed Bi2Te3 regions are highlighted.

In conclusion, we have performed the first STM study of FGT thin films down to the monolayer limit, and characterized the interface with MBE-grown $Bi_2Te_3$ as a platform for coupling 2D magnetism and topological surface states. The observed moiré periodicity is solely due to lattice mismatch, showing that FGT is rotationally aligned to the $Bi_2Te_3$. Tunneling spectroscopy indicates that nearly identical electronic density of states for 1 QL and 2 QL FGT, which are similar to previous reports for bulk FGT as well. This indicates that there is little change in FGT states with thickness. Fe- and FeGe-terminated regions are also accessible in these thin film samples, which provide insight into the layer-resolved density of states. While no clear signature of the FGT magnetism was observed in the SP-STM imaging, our MCD measurements confirm the films are ferromagnetic, and have a higher $T_C$ than exfoliated monolayer flakes. This could be due to the $Bi_2Te_3$ substrate in our study; future work with higher temperatures closer to $T_C$ and in higher magnetic fields would further explore this possible effect.

**Supplemental Information:** RHEED patterns during different phases of growth; Topographic STM images of the control $Bi_2Te_3$ sample.

**Acknowledgements:** This work was primarily supported by the Department of Energy (DOE) Basic Energy Sciences under Grant No. DE-SC0016379. Additional macroscopic characterization of samples was supported by AFOSR MURI 2D MAGIC Grant No. FA9550-19-1-0390.

**Author Contributions:** B.G. acquired and analyzed STM data and wrote the manuscript, A.J.B. grew samples, W.Z., R.B.C., K.R. performed growth and magnetic characterization, R.K.K. and J.A.G. conceived the project and supervised data acquisition and analysis. All authors reviewed and contributed to the manuscript.